\documentclass[sigconf]{acmart}

\renewcommand\footnotetextcopyrightpermission[1]{} 
\usepackage{booktabs} 

\begin{document}
\title[Digitalization of Swedish Government Agencies]{Digitalization of Swedish Government Agencies -- A Perspective Through the Lens of a Software Development Census}
\subtitle{Preprint of paper accepted for the 40th International Conference on Software Engineering,\\May 27--3 June 2018, Gothenburg, Sweden, Software Engineering in Society Track}


\author{Markus Borg}
\orcid{1234-5678-9012}
\affiliation{%
 \institution{RISE SICS AB}
 \streetaddress{P.O. Box 1212}
 \city{Lund} 
 \postcode{43017-6221}
	\country{Sweden}
}
\email{markus.borg@ri.se}

\author{Thomas Olsson}
\affiliation{%
 \institution{RISE SICS AB}
 \streetaddress{P.O. Box 1212}
 \city{Lund} 
 \postcode{43017-6221}
	\country{Sweden}
}
\email{thomas.olsson@ri.se}

\author{Ulrik Franke}
\affiliation{%
 \institution{RISE SICS AB}
 \streetaddress{P.O. Box 1212}
 \city{Stockholm} 
 \postcode{43017-6221}
	\country{Sweden}
}
\email{ulrik.franke@ri.se}

\author{Sa\"id Assar} 
\affiliation{%
\institution{IMT Business School}
\streetaddress{}
\city{Evry} 
\country{France}}
\email{said.assar@telecom-em.eu}

\renewcommand{\shortauthors}{M. Borg et al.}

\begin{abstract}
Software engineering is at the core of the digitalization of society. Ill-informed decisions can have major consequences, as made evident in the 2017 government crisis in Sweden, originating in a data breach caused by an outsourcing deal made by the Swedish Transport Agency. Many Government Agencies (GovAgs) in Sweden are rapidly undergoing a digital transition, thus it is important to overview how widespread, and mature, software development is in this part of the public sector. We present a software development census of Swedish GovAgs, complemented by document analysis and a survey. We show that 39.2\% of the GovAgs develop software internally, some matching the number of developers in large companies. Our findings suggest that the development largely resembles private sector counterparts, and that established best practices are implemented. Still, we identify improvement potential in the areas of strategic sourcing, openness, collaboration across GovAgs, and quality requirements. The Swedish Government has announced the establishment of a new digitalization agency next year, and our hope is that the software engineering community will contribute its expertise with a clear voice. 
\end{abstract}

%
%




\maketitle

\section{Introduction}
With the digitalization of society, more and more organizations become software intensive. Many companies must change to stay competitive on the market, but also public organizations need to improve their software maturity~\cite{fitzgerald_scaling_2017}. Government Agencies (GovAgs) are one type of public sector organizations that inevitably must adapt to the new digital era. While increased software capabilities require considerable investments, funded by tax money, successful scaling of software in GovAgs could also provide multiple benefits. According to the OECD, digital governments can lead to ``more open, transparent, innovative, participatory, and trustworthy governments''~\cite{oecd_council_oecd_2014}. Furthermore, digital governments enable use of technology for creating a new generation of cost-efficient, interactive, and ubiquitous ICT-enabled public services~\cite{gronlund_introducing_2005}.


Digitalization is acknowledged in the Swedish Government's strategy on digital transformation launched in 2017, with an objective to ``become the world leader in harnessing the opportunities of digital transformation''\footnote{http://www.government.se/press-releases/2017/06/action-on-digital-transformation/}. Sweden has a good track record in digitalization in general, e.g., Sweden was ranked third in the World Economic Forum's Networked Readiness Index 2016~\cite{baller_global_2016}, and third in the EU Digital Economy \& Society Index 2017~\cite{european_commission_europes_2017}. However, the digitalization of GovAgs has just begun, and reaping the full benefits of digitization is a challenge for any organization~\cite{fitzgerald_scaling_2017}. 


Software development in the public sector faces unique challenges. GovAgs often have needs unlike any other actor, and when procuring specific IT services a GovAg might finds itself a single buyer on the market. Moreover, sometimes only a few providers on the market offer services geared specifically towards the public sector, thus limiting procurement and sourcing options~\cite{lin_issues_2007,badampudi_decision-making_2017}. Previous work on Swedish GovAgs reports that both software scaling and general software process improvement have been ongoing in recent years~\cite{larsson_revisiting_2014,beskow_scalare_2016}, but there is no overview available of the current state-of-practice in the GovAgs' software projects.

GovAgs' IT is currently a hot topic in the Swedish public debate. The background is that the Swedish Transport Agency outsourced its IT to IBM in 2015. In the summer of 2017, it was revealed that as part of this procurement, the director general of the Swedish Transport Agency had authorized deviations from the applicable laws on information security. Following an investigation by the Swedish Security Service, the director general was fined and dismissed from office\footnote{https://transportstyrelsen.se/en/About-us/statement-about-the-information-in-media-regarding-our-it-public-procurement/}. As the story of insufficient and illegal information security practices at the GovAg made it into the media, questions were raised about how the government had managed the crisis. The questions about who knew what at which time, and which precautionary measures were, or were not taken, have forced two ministers and one state secretary to resign. A third minister was subject to a vote of no-confidence in the parliament, but remained in office as the vote did not gain a majority.


The overall goal of this study is to gage the digitalization of Swedish GovAgs from a software engineering perspective. We contribute a unique view of the permeation of software development in GovAgs -- as an indicator of the digitalization of society. The purpose of this article is not to relate all the details of the recent IT scandal, but it serves as a timely illustration of the importance of informed  decisions. Consequently, a secondary goal of our work is to explore the current state of software development practice in Swedish GovAgs. We present a census of the 240 GovAgs in Sweden, designed to answer the following Research Question (RQ):

\begin{itemize}
\item RQ1. How many government agencies in Sweden develop software to support their core operations?
\end{itemize}
Among the GovAgs identified in RQ1, we continue by exploring:
\begin{itemize}
\item RQ2. What is the distribution of software sourcing strategies?
\item RQ3. Are common software engineering best practices used?
\item RQ4. How are software qualities addressed? 
\end{itemize}


The rest of the paper is organized as follows: Section~\ref{sec:bg} defines digitalization and introduces related work on e-government and software engineering in the public sector. Section~\ref{sec:method} describes our data collection and analysis. In Section~\ref{sec:res} we present our results and discuss our findings in the light of the RQs. Finally, Section~\ref{sec:conc} summarizes the paper, puts it in the context of digitalization in Sweden, and outlines future work.

\section{Background and Related Work} \label{sec:bg}
In this section, we present background knowledge about digitalization and e-government. We also present related work on software development in the public sector, in Sweden and abroad.

\subsection{Digitalization of Society and E-Government}
Digitization is usually defined as ``the integration of digital technologies into everyday life by the digitization of everything that can be digitized''~\cite{ochs_it_2018}.
This profound change in the way business is conducted is known as ``digital transformation''~\cite{matt_digital_2015} or ``digitalization''. Digitalization implies disrupting organizational structures and adopting new innovative perspectives for the definition of commercial products and the creation of business value~\cite{schmidt_digitization_2015}.

In the public sector, digitization and digitalization are generally considered as extensions of e-government. Although e-government was initially considered as a particular form of e-commerce consisting in providing online documents and services to citizens~\cite{gronlund_introducing_2005}, its scope is much wider and includes political goals such as institutional reforms, government modernization, and the introduction of new democratic practices~\cite{assar_back_2011}. Denmark, another country on the Scandinavian Peninsula, established an Agency for Digitisation in 2011, with the mission to ``speed up the digitalization processes required to modernize the Danish welfare society''.


The evaluation of e-government endeavors has received much research attention~\cite{gupta_e-government_2003,luna-reyes_towards_2012}. There are multiple criteria for evaluation, e.g., quality of information, level of personalization, level of digitization, and the scope of online services. The United Nations e-government readiness index, a large and evolving set of factors, has shifted focus from e-inclusion in 2005, to financial leverage in time of the 2010 crisis, to support of sustainable development in 2016~\cite{united_nations_department_of_economic_and_social_affairs_united_2005}. This shift of focus implies that a country's position can change a lot from one year to another. For example, Sweden was in third place in 2005, then dropped from the top 10 in 2010, and is ranked sixth in the 2016 UN e-government readiness index~\cite{united_nations_department_of_economic_and_social_affairs_united_2005}. 

Bridging e-government and the next subsection, there is a movement to make public \emph{data} as well as \emph{software} developed in publicly funded projects open~\cite{free_software_foundation_europe_fsfe_public_2017}, in Sweden and in other countries. Both the European Commission\footnote{Open data https://ec.europa.eu/digital-single-market/en/open-data} and the Swedish Government\footnote{http://www.regeringen.se/pressmeddelanden/2016/07/regeringen-oppnar-dorren-for-mer-oppen-data} argue that more open data has the potential to lead to new innovations that address societal challenges, as well as increased transparency of governments.

\subsection{Software Engineering in the Public Sector}
As software is the founding element for all ICT based public services, the development and acquisition of software together with the optimization of their information systems architectures and operations have acquired strategic importance for all government activities. Moreover, the spectacular failure of some large public information systems projects, e.g., the Human Resource and Payroll system in France in 2014~\cite{moreaux_logiciel_2014}, have led public authorities to strongly enforce best practices and modern approaches to software development projects. For example, The Swedish Tax Agency recently announced a transition to agile development methods~\cite{lindstrom_nu_2017} and information security issues gets increasing attention by the Swedish Government\footnote{http://www.regeringen.se/pressmeddelanden/2017/08/sakrare-statlig-it-drift-genom-okad-samordning/}.

Some research claims that public sector projects tend to be poorly conducted compared to its private sector counterparts~\cite{goldfinch_pessimism_2007}. On the other hand, in a study conducted in Norway in 2012, a number of project indicators were compared between public and private organizations, and the authors found no significant differences~\cite{krogstie_comparing_2012}. Nonetheless, there are indeed contextual differences between software projects in the public and private sectors. Rosacker and Rosacker report three major differences~\cite{rosacker_information_2010}: 1) the private sector is market-driven, whereas the public sector has less competition, 2) managers in the private sectors are primarily accountable to immediate customers and shareholders, in the public sector the stakeholders represent a broader group of constituents, and 3) public organizations can be subject to more forceful laws and regulations than private companies. 

We identified some previous work on software projects in Swedish GovAgs. Larsson and Borg compared aspects of software engineering in large companies with its counterpart in a Swedish GovAg~\cite{larsson_revisiting_2014}. They report that a particular challenge for the GovAg is that their goals depend on external directives that might change as political powers shift, either on the national or European Union (EU) level. Another difference is that the GovAg does not develop a solution for an open market, instead they implement a solution required by the EU commission -- with substantial financial penalties if they are not fulfilled on time. The same authors later presented another study at the same Swedish GovAg~\cite{larsson_testing_2016}, focusing on testing of QRs, but no findings appear to be unique to the public sector. 

Looking beyond Sweden's borders, Bannerman reported that software risk management in Australian GovAgs was unsystematic or informal -- despite very large investments~\cite{bannerman_software_2007}. Furthermore, Bannerman reported that rigid plan-driven waterfall development dominated, thus offering limited possibilities to adapt to risks during project execution. Patanakul presented a study of problems in 14 large (>\$1B) public sector projects in the US, UK, and Australia~\cite{patanakul_managing_2014}. His results corroborate Bannerman's conclusion that risk management is inadequate, and also highlight that unclear requirements are a common cause for failed projects. Ziemba and Kolasa also addressed software risk management, but in the Polish public sector~\cite{ziemba_risk_2015}. In line with Larsson and Borg~\cite{larsson_revisiting_2014}, they argued that the public sector context introduces additional challenges due to an increased sensitivity to changing external directives, i.e., changing government processes or legal regulatory frameworks.


\section{Method} \label{sec:method}
We conducted a census of software development at Swedish GovAgs, i.e., an inquiry to gather information about every individual in a population~\cite{fowler_survey_2013}. The target population was all GovAgs listed by Statistics Sweden\footnote{http://www.scb.se/myndighetsregistret} on January 1, 2017, in total 240 GovAgs. The data collection was based on the Swedish Freedom of the Press Act~\cite{swedish_ministry_of_justice_public_2009}, stating that anyone is entitled to read the documents held by public authorities. 

\begin{figure}
\includegraphics[scale=0.60]{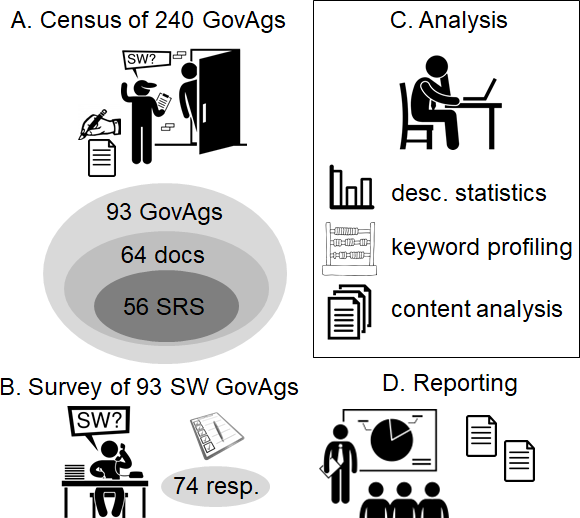}
\caption{Overview of the research design: a census followed by a survey and an analysis based on triangulation. The reporting consists of this paper and a technical report~\cite{borg_digitalization_2018}.}
\label{fig:process}
\end{figure}

Figure~\ref{fig:process} shows an overview of the research design. We initiated the data collection process by sending a formal request per email to all 240 GovAgs at 12 AM on January 1, 2017 (cf., A. in Fig.~\ref{fig:process}). With support from an archivist, we formally requested a sample of software development documentation, with an emphasis on System Requirements Specifications (SRS), from an ongoing or completed software project. We stressed that the project should be related to the GovAg's core operations and go beyond web content management. We distributed up to three reminders with two-month intervals, and eventually 93 GovAgs confirmed that software development occurs. As some GovAgs referred to secrecy, we obtained documents from 64 GovAgs, and 56 GovAgs provided at least one SRS.  



We complemented our census with a questionnaire-based survey of the 93 GovAgs that confirmed software development (cf. B. in Fig.~\ref{fig:process}). The questionnaire (available in the accompanying technical report~\cite{borg_digitalization_2018}), distributed in mid-August 2017 to contacts identified during the census, consisted of closed questions. A majority of the questions consisted of three Likert scales, addressing: 1) development process, 2) sourcing strategies, and 3) development context, in total encompassing 25 Likert items. The selection of statements was influenced by SWEBOK v. 3.0~\cite{ieee_computer_society_guide_2015}, ISO/IEC 25010 on quality requirements~\cite{international_organization_for_standardization_systems_2011}, Murhphy-Hill \textit{et al.}'s comparison between game development and traditional software engineering~\cite{murphy-hill_cowboys_2014}, and Badampudi \textit{et al}'s research on sourcing options~\cite{badampudi_decision-making_2017}. We calculated descriptive statistics for the survey, including Spearman's rank-order correlation coefficients.

We collected data of three types: 1) direct correspondence between GovAgs' officials and the first author per mail and/or telephone, 2) obtained official documents from 64 GovAgs, and 3) 74 survey responses. The first author cataloged the correspondence in a spreadsheet. All written communication was saved, and telephone calls (typically resembling interview sessions) were documented in notes. Our research design enables data and method triangulation~\cite{robson_real_2002}, i.e., we draw conclusions based on multiple sources of evidence collected using different methods. We describe the document analysis and the main threats to validity next.


\subsection{Document analysis} \label{sec:docs}
First, we performed keyword frequency profiling~\cite{rayson_comparing_2000} (cf. C. in Fig.~\ref{fig:process}), or rather keyword presence profiling, to determine which software qualities defined in ISO/IEC 25010 (from now on: keywords) occur in the GovAgs' documents. The goal of this light-weight analysis was to identify whether concepts relating to software qualities occur in the SRSs. Second, we performed qualitative content analysis~\cite{bowen_document_2009} of the obtained documents, predominately using predefined topics and codes~\cite{robson_real_2002}.


The keyword profiling evolved during the process. We validated the keywords by assessing the resulting frequency profiles and by manually investigating a subset by reading keywords in the context of the SRSs. As a result, some keywords required special treatment. ``Security'' and its Swedish translation ``s\"akerhet'' are too general concepts that also are used in everyday expressions. On top of that, ``s\"akerhet'' is used to denote both safety and security in Swedish. Hence, we decided to refine the security concept into: 1)~confidentiality, 2)~integrity, and 3)~availability -- motivated by the fact that Swedish GovAgs are mandated to classify their information assets accordingly\footnote{https://www.msb.se/externdata/rs/94a3d208-2ac4-48a1-84f2-208268f5767e.pdf}. Moreover, the Swedish translations of ``availability'' (``tillg\"anglighet'', also meaning ``accessability'') and ``reliability'' (``tillf\"orlitlighet'', also used for trust in general) suffer from similar vagueness problems, thus these keywords were restricted to English. For the keyword ``confidentiality'', we use the two keywords ``konfidentialitet'' and ``sekretess'' and for ``portability'' we used both ``portabilitet'' and ``flyttbarhet'' in Swedish to capture all relevant occurrences of the concepts. All translations are available in the accompanying technical report~\cite{borg_digitalization_2018}.

Content analysis in documents is often used in combination with other research methods as a means of triangulation~\cite{bowen_document_2009}. We used content analysis for this purpose mainly in relation to RQ3 and RQ4, i.e., data triangulation to identify evidence, or at least indications, of implemented best practice presence and quality foci. Due to the large amounts of documents obtained from the GovAgs, a comprehensive document analysis was not within the scope of this study. Consequently, the content analysis can only be used to draw conclusions on the presence of phenomena, and not absence.

\subsection{Threats to Validity} \label{sec:threats}
This section presents the main threats to the validity of our findings, organized in three parts. \textit{External validity} concerns the generalizability of our results. First of all, the census part of our study does not suffer from issues related the sampling and response rate. We sent formal requests to the entire population of Swedish GovAgs and everyone was required to answer, thus we consider it highly reliable. Through the census, we collected firsthand contacts with government officials responsible for software development, which we later used to reach a very high response rate (76.6\%) in the survey part. As our respondents are GovAgs and not individuals, we respond to the call by Stavru that more surveys should target organizations instead of personal opinions~\cite{stavru_critical_2014}. That said, we cannot be sure that the entire GovAgs is behind the individual answers. 

\textit{Content validity} concerns how much a measure represents every single element of a construct. For the survey, we had to limit the number of Likert items to keep the response times reasonable. We selected statements based on related work, incl. SWEBOK~\cite{ieee_computer_society_guide_2015}, but obviously more Likert items would have captured additional aspects of software development. Understanding software development requires more than a list of closed questions, thus we plan for future work with in-depth case studies at a selection of GovAgs.

Finally, \textit{construct validity} refers to how an operational definition of a variable actually reflects the true theoretical meaning of a concept. The major threat to our study is whether our inquiry about software development was properly interpreted by the respondents. Many GovAgs appeared to lump together all matters related to IT, not distinguishing software development from general IT operations. However, we believe that the document analysis filtered out any GovAgs that do not develop software internally.
The keyword profiling also introduces threats to construct validity, as the approach cannot completely capture all concepts of software qualities. First, there might be false positives, i.e., wrongly identifying a quality focus as present in an SRS. Second, we could have false negatives, i.e., failing to identify a quality focus that indeed was described in an SRS. We mitigated both threats by manual assessment of a sample of both keywords and SRSs.

\section{Results and discussion} \label{sec:res}
All GovAgs in Sweden answered the census, i.e., whether software development related to the GovAg's core operations occurs.
Seventy-four of the 93 GovAgs that confirmed software development answered our survey (cf. Table~\ref{tab:survey}), i.e., a response rate of 79.6\%. Five GovAgs could not answer the survey for confidentiality reasons and three GovAgs declined to answer. We are still expecting answers from two GovAgs, but the remaining 9 GovAgs did not reply to our emails. This section synthesizes findings from the census, survey, and document analysis to answer the RQs. 

\subsection{Software development at Swedish Government Agencies -- A Census (RQ1)} \label{sec:res1}
All GovAgs in Sweden answered the census, but three GovAgs were discontinued during the execution of the study, resulting in a total population of 237 GovAgs. Ninety-three out of 237 GovAgs (39.2\%) answer that they have in-house software development, performed either with employed developers or contractors. However, several GovAgs answer that another GovAg provides custom software solutions for them, e.g., all 21 County Administrative Boards have centralized the IT development organization to the agency in V\"astra G\"otalands L\"an. Furthermore, seven GovAgs report that their respective parent GovAgs provide all software, e.g., the National Board of Health and Welfare provides the smaller National Medical Responsibility Board with all software solutions.

Most GovAgs disclosed documentation from in-house software projects. We obtained software documentation from 64 out of 93 GovAgs (68.8\%). Nine GovAgs answered that the documentation was secret, e.g., the Prosecution Authority, the Coast Guard, the Election Authority, and the Armed Forces. By the time of this writing, we are still waiting for documentation from 20 GovAgs, including some large GovAgs such as the National Property Board.

The census also revealed that several GovAgs have considerable software engineering know-how despite having no in-house development. Sixteen of the 237 GovAgs report that they develop software and systems requirements in-house, which then are used for public procurement. We obtained SRSs from all these 16 GovAgs, and conclude that software engineering knowledge exists also in GovAgs not covered among the 93 that have internal development. Considerable requirements engineering skills are needed when specifying large systems in the public sector, and several GovAgs without internal development do it well -- however, none of these 16 GovAgs are part of the analysis in the remainder of the paper.

Table~\ref{tab:volume} shows an overview of the volume of software development conducted at Swedish GovAgs. The most common number of developers is between 5 and 19, but several GovAgs report having large development organizations -- some with more than 100 software developers. Note that GovAgs such as the Armed Forces refer to secrecy, and that we are still waiting for figures from some large GovAgs, e.g., the Public Employment Agency and Statistics Sweden, thus the number of large development organizations is likely to be even higher.

\begin{table}[]
\centering
\caption{Number of software developers at Swedish GovAgs and the proportion of employed development resources.}
\label{tab:volume}
\begin{tabular}{|l|cl|l|c}
\cline{1-1} \cline{4-4}
\multicolumn{1}{|c|}{\begin{tabular}[c]{@{}c@{}}\#Developers\end{tabular}} &  &  & \multicolumn{1}{c|}{\%Employed} &  \\ \cline{1-2} \cline{4-5} 
1-4 & \multicolumn{1}{c|}{9} &  & 0-20\% & \multicolumn{1}{c|}{10} \\ \cline{1-2} \cline{4-5} 
5-19 & \multicolumn{1}{c|}{30} &  & 21-40\% & \multicolumn{1}{c|}{11} \\ \cline{1-2} \cline{4-5} 
20-49 & \multicolumn{1}{c|}{13} &  & 41-60\% & \multicolumn{1}{c|}{15} \\ \cline{1-2} \cline{4-5} 
50-99 & \multicolumn{1}{c|}{8} &  & 61-80\% & \multicolumn{1}{c|}{24} \\ \cline{1-2} \cline{4-5} 
100-199 & \multicolumn{1}{c|}{8} &  & 81-100\% & \multicolumn{1}{c|}{11} \\ \cline{1-2} \cline{4-5} 
\textgreater199 & \multicolumn{1}{c|}{6} &  & Missing & \multicolumn{1}{c|}{3} \\ \cline{1-2} \cline{4-5} 
\end{tabular}
\end{table}

Table~\ref{tab:volume} also shows an overview of the proportion of employed development resources, as opposed to contractors, in the GovAgs' development organizations. The proportion varies from no employed software developers at all (e.g., the Energy Markets Inspectorate and the Environmental Protection Agency) to having nothing but employed developers (e.g., the Swedish University of Agricultural Sciences and the National Veterinary Institute). Typically, development at Swedish GovAgs involves a mix between in-house resources and contractors, i.e., most often between 40-80\% of the developers are employed by the GovAg. These figures apply to all respondents with 50 or more developers, except the Prison and Probation Service and the Council for Higher Education (70\% contractors each) and the Meteorological and Hydrological Institute (85\% employed developers). 

Table~\ref{tab:survey} shows the responses to the three Likert scales collected as part of the survey. In the rest of the paper, we refer to individual statements using their IDs in bold font, e.g., ``\textbf{S3d}''. Based on the survey, we continue by providing a general overview of software development at Swedish GovAgs.

Agile development is an important contextual factor when describing contemporary software engineering, thus it receives particular attention in our study. A clear majority of the GovAgs (65 out of 73, 87.8\%) report that their software development is more agile than plan-driven (\textbf{S3b}). Four other statements are also related to agile development, and can thus be used to corroborate the level of agility. Lean software development principles (\textbf{S3c}, 70.3\%) are common among the GovAgs, and a majority also confirm implementing DevOps (\textbf{S3d}, 73.0\%). Note, however, that Lwakatare \textit{et al}. recently showed that practitioners have different interpretations of what constitutes DevOps~\cite{lwakatare_exploratory_2016}. Finally, continuous integration (\textbf{S3m}) is used in 68.9\% (46 out of 74) of the GovAgs, but test automation (\textbf{S3n}) is not a standard practice; only 32 out of 74 (43.2\%) agree or strongly agree to the statement.

\begin{table*}
\centering
\caption{Results from the survey. The Likert distributions show ``Strongly disagree'' to the left and ``Strongly agree'' to the right. }
\label{tab:survey}
\includegraphics[scale=0.5]{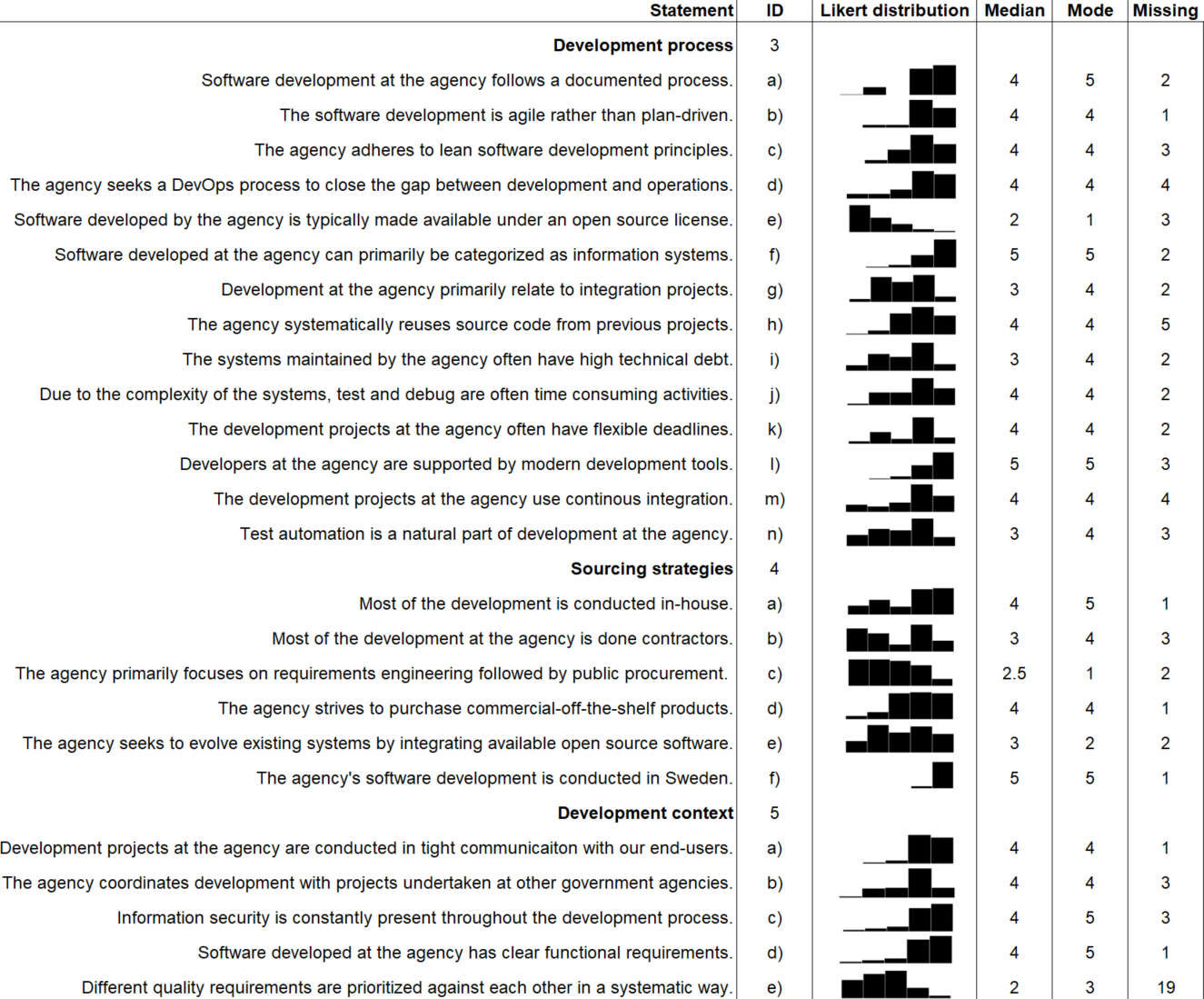}
\end{table*}

Analyzing the five statements related to agile, we find that many GovAgs get high agility ratings in our study. Most of them are small, including several universities, but also several large GovAgs appear to adhere to agile development practices, e.g., the Board of Agriculture and the Swedish Customs. We identified evidence of agile practices in the document analysis, such as Scrum backlogs from Linnaeus University and the Companies Registration Office, but also examples of traditional plan-driven documents with agile elements such as user stories. Our finding is in line with the dominance of agile and iterative methods reported by Scott \textit{et al.} in the HELENA survey~\cite{scott_initial_2017}, i.e., Scrum, Kanban, and DevOps are all more popular than waterfall processes, both in Sweden and worldwide.  

A recent example of the agile movement influencing Swedish GovAgs is that the Tax Agency, one of the most software-intensive GovAgs, announced a transition to agile methods in September 2017~\cite{lindstrom_nu_2017}. 
The level of agility we report for Swedish GovAgs contrasts with findings by Bannerman from 2007, who reported that Australian GovAgs were mainly plan-driven~\cite{bannerman_software_2007}. 

A rationale for GovAgs to use agile methods is that a majority (45 out of 75, 60.8\%) report having flexible deadlines (\textbf{S3k}). This suggests that it might be more important for a GovAg to be responsive to change rather than to follow a traditional plan, in line with the Agile Manifesto. The reported flexibility was unexpected, as previous work rather emphasized that software projects in the public sector often are subject to legislative changes with major consequences if not implemented in time~\cite{larsson_revisiting_2014}.


We designed three questions geared at gaging the nature of the products and services developed by the GovAgs. Sixty-five out of 74 (87.8\%) answer that the agency mainly develops information systems (\textbf{S3f}), i.e., systems intended to store, manage, and present information. The respondents are split in two groups regarding the statement on integration projects (\textbf{S3g}), such as described by Larsson \textit{et al.}~\cite{larsson_testing_2016}: 29 GovAgs agree and 25 disagree. 
Finally, a majority of the respondents (44 out of 74, 59.5\%) answer that test and debug are time consuming activities due to the complexity of the systems (\textbf{S3j}). Thus, our results indicate that the systems developed in the public sector are far from trivial, and in comparison to Murphy-Hill \textit{et al}'s study on development at Microsoft~\cite{murphy-hill_cowboys_2014}, they could be even more complex than private sector counterparts. 

\subsection{Sourcing strategies (RQ2)} \label{sec:res2}
A recurring strategic consideration in software projects is whether to develop an asset internally or acquire it from external sources. In software projects, the decision involves choosing between several different sourcing options~\cite{badampudi_decision-making_2017}. First, a GovAg can develop a software asset in-house. Second, software can be acquired externally by 1) specifying requirements and outsourcing of development through public procurement, 2) purchasing commercial-off-the-shelf (COTS) software through public procurement, and 3) acquiring existing open source software (OSS). Note that previous research on strategic software sourcing mainly focused on market-driven contexts in the private sector; public-sector organizations are primarily designed to be ``fair, open, objective, and accountable'' rather than efficient~\cite{cilek_hedonic_2004}.

Most GovAgs respond that a majority of the software development is conducted by employed resources (45 out of 74, 60.8\%), while only 21 disagree with the corresponding statement (\textbf{S4a}) (28.4\%). On the other hand, roughly the same proportion of respondents agree and disagree with the statement that most development is conducted by contractors (\textbf{S4b}, 31 vs. 34 out of 74). Thus, as the statements \textbf{S4a} and \textbf{S4b} were designed to expose an inverse relationship, the results are contradicting. We believe that a fraction of the respondents considered contractors working on the GovAgs premises as in-house resources, possibly due to long contracts. Regarding development abroad, currently a highly sensitive topic in the Swedish press, all respondents (but a missing answer) report that the development is conducted in Sweden: 67 (90.5\%) even strongly agree to the statement (\textbf{S4f}). Regarding operations, however, Ottsjo and Kristensson conducted a survey after the Transport Agency scandal revealing that 11 Swedish GovAgs manage data in other countries~\cite{ottsjo_myndigheternas_2017}, i.e., the solution that caused the 2017 crisis.

Our survey identifies a handful of relevant correlations between a focus on contractors (\textbf{S4f}) and the characteristics of software projects. The following statements are all correlated with \textbf{S4b} (statistically significant moderate correlations, $0.31\leq|\rho|\leq0.44$). Software development primarily conducted by contractors is correlated with: 1) high technical debt (\textbf{S3i}, $\rho=0.31$) and negatively correlated with 2) adherence to lean principles (\textbf{S3c}, $\rho=-0.44$), 3) security awareness throughout development (\textbf{S5c},  $\rho=-0.36$), 4) releasing source code as OSS (\textbf{S3e}, $\rho=-0.33$), and 5) coordinated development across GovAgs (\textbf{S5b}, $\rho=-0.31$). 

While we cannot make causal claims, our findings suggest that contractors and technical debt co-occur in software projects at Swedish GovAgs. We found no previous work on the connection between technical debt and contractors in the public sector, the closest result instead targeting outsourcing. Assuncao \textit{et al.} stressed the need for Brazilian GovAgs to check the quality of source code developed by third parties~\cite{assuncao_technical_2015}. Lin \textit{et al.} concluded that outsourcing in the public sector threatens the organizational memory, which we believe is fundamental to avoid technical debt. Another correlation suggests that GovAgs with a high proportion of contractors focus less on security. A noteworthy exception is The National Board of Institutional Care, a GovAg with a small development organizations consisting almost exclusively of contractors, that ranks high on security focus and awareness despite the external resources. 

Nineteen out of 74 respondents claim that the GovAg mainly does requirements engineering followed by public procurement (\textbf{S4c}, 25.7\%), most of them having fewer than 20 developers. Exceptions include the Customs Service and the Government Offices of Sweden, both having 50-99 developers but still focusing on procurement of development effort. Lin \textit{et al.} reported in 2007 that 92.3\% of the public sector organizations in Australia outsourced at least parts of their IT functions~\cite{lin_issues_2007}. They found a negative correlation between the percentage of outsourcing and organization size, i.e., a result in line with our findings, although our focus is on software development rather than IT functions. Finally, a majority of the GovAgs has an ambition to buy COTS systems rather than developing their own solutions (\textbf{S4d}, 58.1\%), only nine GovAgs disagree with the statement. 

Few GovAgs develop a majority of their software under open source licenses (6 out of 74, 8.1\%). Instead, as many as 54 out of 74 (73.0\%) disagree to the corresponding statement (\textbf{S3e}). This is interesting in the light of calls for more open data and open software in the public sector. It may be that even though the intention is to go in the direction of openness, the pace is limited by technical and organizational inertia. This hypothesis gains some support from the fact that while only six of the responding Swedish GovAgs predominantly develop software under open source licenses, considerably more GovAgs strive to acquire OSS solutions when investing in \emph{new} systems (\textbf{S4e}, 31 out of 74, 41.9\%). If this observation is indeed an indicator of an ongoing development, it seems likely that we will see more open source software in the public sector in the future.

Many GovAgs develop and maintain software solutions offering similar features. The potential for commodity components and services became evident as our understanding grew during the study. Statement \textbf{S5b} reflects this, as 46 out of 74 GovAgs (62.2\%) respond that they align development efforts with other GovAgs, e.g., through knowledge exchange activities and shared solutions. There is still a potential for improvement, however, as 13 GovAgs do not collaborate much with others. In addition, only half of the GovAgs report a systematic approach to source code reuse (\textbf{S3h}, see Section~\ref{sec:res3}) -- an approach that could facilitate collaboration across GovAgs.

There are current initiatives to improve GovAg collaboration, at least on a strategic level. In February 2017, Statens servicecenter, a GovAg supporting GovAg efficiency, concluded that a common cloud solution for Swedish GovAgs could reduce costs by 30\%~\cite{statens_servicecenter_en_2017}. Similar initiatives have been reported also in Korea~\cite{lee_implementing_2013} and China~\cite{liang_integrating_2013}. In August, in the aftermath of the IT scandal, the Swedish Government commissioned the Social Insurance Agency to host a secure cloud solution for other GovAgs\footnote{https://computersweden.idg.se/2.2683/1.687526/it-drift-forsakringskassan}. The Social Insurance Agency is among the most digitally mature GovAgs in Sweden, and enabling both shared technical solutions and knowledge transfer appears promising. However, as a focus on contractors is negatively correlated with cross-GovAg collaboration, our recommendation for Swedish GovAgs is to maintain an employee vs. contractor ratio that still ensures sufficient internal competency.

\subsection{Software engineering best practices (RQ3)} \label{sec:res3}
The survey encompasses a self-assessment for GovAgs in relation to a selection of acknowledged software engineering best practices. Inspired by SWEBOK~\cite{ieee_computer_society_guide_2015}, we designed a number of Likert items that reflect generally accepted practices.

\textbf{Requirements engineering} is known as a foundation for software quality, and poor requirements have turned many software projects into failures~\cite{ieee_computer_society_guide_2015}. Sixty-four out of 74 GovAgs (86.5\%) report that the software development is guided by clear functional requirements (\textbf{S5d}). The document analysis of the 56 GovAgs that provided SRS partly supports this -- the requirements specification practices appear generally mature. For example, The Board of Student Finance and Swedish University of Agricultural Sciences provided structured text requirements with unique identifiers and combine them with allocation to development sprints. Several GovAgs, e.g., the Swedish ESF Council and the Environmental Protection Agency, use textual use case descriptions, sometimes in combination with UML diagrams. 

The finding that software development in GovAgs is supported by  clear functional requirements corroborates and generalizes results by Larsson and Borg, who reported that one specific large Swedish GovAg had well-defined and verifiable functional requirements~\cite{larsson_revisiting_2014}. It had not always been that way though, as the authors reported that the requirements engineering practices had matured considerably in recent years. On the contrary, Patankul reported that unclear requirements often plague software projects in the public sector~\cite{patanakul_managing_2014}. In the same vein, Mendez-Fernandez \textit{et al.} report in the NaPiRE survey that the incomplete and/or hidden requirements is a major challenge to all types of software projects, no matter size or development agility~\cite{mendez_fernandez_naming_2017}. Thus, we see a need for further studies on the nature of the functional requirements in Swedish GovAgs -- how do they manage to capture them with such precision?

\textbf{Software processes}, no matter if they are agile or plan-driven, facilitate communication and coordination, and support development of high-quality software products\cite{ieee_computer_society_guide_2015}.
Most GovAgs (62 out of 74, 83.8\%) agree that their development is guided by documented processes (\textbf{S3a}). Note that while the first generation of agile methods (as captured in the Agile Manifesto) downplays the role of processes, later agile constructs such as Scrum and DevOps have well-defined processes. As a majority of the GovAgs consider themselves agile, it appears likely that more recent agile processes are implemented, or possibly hybrid approaches~\cite{scott_initial_2017}.

\textbf{Software reuse} is recognized as a key factor in improving productivity and competitiveness~\cite{ieee_computer_society_guide_2015}, but it requires strategic vision and supporting processes. Roughly half of the GovAgs report a systematic approach to source code reuse (\textbf{S3h}). Our results identify software reuse as an improvement potential, possibly in combination with collaboration across GovAgs as discussed in Section~\ref{sec:res2}.

Regarding development \textbf{tool support}, as many as 86.5\% (64 out of 74) agree that appropriate tool support is available (\textbf{S3l}) -- a rather high number given that surveys often indicate that practitioners call for better tools. Consequently, it appears that Swedish GovAgs are not obstructed by a lack of modern development tools.

GovAgs are not spared from challenging legacies and short-termed solutions. Thirty-five out of 74 GovAgs (47.3\%) report that their systems have considerable \textbf{technical debt} (\textbf{S3i}). Assuncao \textit{et al.} claims that the issue of technical debt is increasing in the public sector, and investigated the phenomenon in software projects run by the Brazilian Ministry of Communications~\cite{assuncao_technical_2015}. Whether the technical debt is increasing in Swedish GovAgs we cannot say, but it is clearly an issue that should be considered -- perhaps GovAgs should avoid hiring too many contractors as discussed in Section~\ref{sec:res2}. 

We regard two statements used in the agility assessment in Section~\ref{sec:res1} to be relevant also as software engineering best practices. Both are widespread approaches to limit negative consequences of big bang integration, namely \textbf{continuous integration} (\textbf{S3m}) and \textbf{test automation} (\textbf{S3n}). While the practices are complementary in nature, 62.6\% of the GovAgs use the former practice and only 43.2\% the latter.

A successful software development project requires interactions with stakeholders, in particular \textbf{feedback from end-users}~\cite{ieee_computer_society_guide_2015}. As many as 67 out of 74 GovAgs (90.5\%) agree that they communicate frequently with the future end-users, and only two single GovAgs disagree with the corresponding statement (\textbf{S5a}). This is a high number, suggesting that software projects in Swedish GovAgs are well connected with end-users such as GovAg officials or the Swedish citizens. Also, collaboration with end-users is a cornerstone in agile development, which a majority of the GovAgs adhere to. Moreover, Larsson and Borg reported that close collaboration between end-users and the software developers was one of six agile practices followed in a large Swedish GovAg~\cite{larsson_revisiting_2014}.

When developing secure software, \textbf{security} must be built into the development process from the start~\cite{ieee_computer_society_guide_2015}. Fifty-nine out of 74 GovAgs (79.7\%) agree that security awareness permeates the entire development process (\textbf{S5c}) (32 GovAgs even strongly agree, 43.2\%). Only six GovAgs disagree with the statement, including three GovAgs involved in the finance sector (the Financial Supervisory Authority, the Enforcement Authority, and the National Government Employee Pensions Board). On the other hand, disagreement might indicate a particular maturity, i.e., these GovAgs might be well aware of the need to build in security from the start, and thus emphasize that even more security awareness would be beneficial to their software projects. 

Finally, we report four correlations between software projects guided by clear functional requirements (\textbf{S5d}) and other statements. The following statements are all correlated with \textbf{S4b} (statistically significant strong or moderate correlations, $0.31\leq\rho\leq0.59$). Software development primarily conducted by contractors is strongly correlated with: 1) following a documented process (\textbf{S3a}, $\rho=0.59$) and moderately correlated with 2) coordination with other GovAgs (\textbf{S5b}, $\rho=0.40$), 3) close communication with end-users (\textbf{S5a}, $\rho=0.36$), and 4) adherence to lean principles (\textbf{S3c}, $\rho=0.31$).

Without claiming any causality, we notice that GovAgs guided by clear functional requirements also report other advantages. Clear requirements go hand in hand with other types of clarity, including processes and communication both with end-users and other GovAgs. Requirements are known to be a vehicle for communication~\cite{ieee_computer_society_guide_2015}, and  properly managed they might support collaboration across GovAgs. The correlation between clear requirements and lean principles is less obvious, but perhaps lean principles force requirements into brief constructs with no wasteful description. Alternatively, clear requirements might help a GovAg to focus on core development activities that bring value.

Our general impression is that the maturity of software development in the public sector resembles the software projects we have studied in the private sector. Most GovAgs with 50 or more developers implement established best practices. Two exceptions, both in the financial sector, are the Financial Supervisory Authority and the National Government Employee Pensions Board -- two anomalies that constitute interesting directions for future work. Nevertheless, our results confirm the conclusion by Krogstie~\cite{krogstie_comparing_2012}, i.e., software development in the public appears as mature as in the private sector.

\subsection{Software qualities (RQ4)} \label{sec:res4}
Quality requirements are a well-known software engineering challenge~\cite{berntsson_svensson_quality_2011}. In the survey, we explicitly asked the respondents to select the three most important qualities as defined by ISO/IEC 25010~\cite{international_organization_for_standardization_systems_2011}, listed in Table~\ref{tab:keywords}. In addition, 53 GovAgs ranked the three qualities selected. In the keyword profiling, we found that 24 out of 56 GovAgs (42.9\%) provided SRSs in which one or more of the keywords occur. 

By far the most commonly selected quality in the survey was \textbf{functional suitability}, listed by 54 out of 74 GovAgs (73.0\%), and also ranked as the most important quality by 29 GovAgs. The results suggest that GovAgs are highly results-oriented, for most respondents it is more important that software provides all functions required, rather than \textit{how} it is provided, i.e., performance, usability, security etc. and the rest of the qualities tend to be secondary considerations. The focus on functional suitability is not surprising, however, but in line with tendencies identified by Sentilles \textit{et al.} in component-based software engineering~\cite{sentilles_property_2016}. On the other hand, the keyword profiling showed that the functional suitability occurs only in an SRS from a single GovAg. Thus, it is evident that other keywords are required to capture this concept.


\textbf{Security} was listed by 46 out of 74 GovAgs (62.2\%), and is also ranked as the most important quality by 12 GovAgs (cf. Table~\ref{tab:keywords}). Moreover, the security related keywords (confidentiality, integrity, and availability) are the most prevalent in the SRS as they occur in SRSs from 20 GovAgs. For some GovAgs, both the survey results and the keyword profiling show that security is central, e.g., the Medical Products Agency and the Land Registration Authority. For others, the obtained SRS did not support the claimed focus on security, including the Tax Agency and the Transport Agency.  In most cases the discrepancy can probably be explained by the limited SRS sample obtained, but it might also indicate that the security focus has not been fully operationalized. 


Roughly half of the respondents list \textbf{usability} (34 out of 74, 45.9\%) and \textbf{reliability} (30 out of 74, 40.5\%) among the top three most important qualities. Seven GovAgs stand out by listing usability as the single most important quality, including the Swedish Police. The keyword profiling showed that usability occurs in SRSs from 16 of the 55 GovAgs, i.e., both the survey and the SRSs indicate its importance. The term reliability was only found in one single SRS, since we could not search for the Swedish translation as described in Section~\ref{sec:docs}.

\textbf{Performance efficiency} appeared more important in the keyword profiling than in the survey. Performance keywords occur in SRSs from 17 out of 56 GovAgs, but performance was only prioritized by 11 out of 74 GovAgs (14.9\%) in the survey -- none of them ranking it as the most important quality. A possible explanation for this is that it is fairly easy to write performance QRs in comparison to the other qualities. However, if the SRSs specify unnecessarily strict performance QRs, the overall development costs might increase. We speculate that there is a lack of strategic direction leading to unnecessary performance QRs. Thus, there might be a potential for GovAgs to improve how quality targets are set, e.g., using the QUPER model proposed by Regnell \textit{et al.}~\cite{regnell_supporting_2008}.

Three of the ISO/IEC 25010 qualities appear to be less important to Swedish GovAgs. \textbf{Maintainability} was listed  among the top three qualities by 16 out of 74 GovAgs (21.6\%), but its keywords occur only in SRSs from three GovAgs. \textbf{Compatibility} and \textbf{portability} were both only listed by a handful of GovAgs each, although the related keywords occur in SRSs from 10 and six GovAgs, respectively. As we argued for performance QRs, this might indicate that compatibility and portability QRs are comparatively straightforward to specify.

Finally, we explored how many GovAgs use systematic approaches to prioritize different QRs, e.g., the analytic hierarchy process or the \$100 test. Only nine out of 74 GovAgs (12.2\%) agree to the statement (\textbf{S5e}) and as many as 28 disagree (37.8\%). However, roughly a quarter of the respondents neither agree nor disagree (18 out of 73, 24.3\%) -- possibly indicating limited understanding of what a systematic method means. This is interesting and also a bit worrying, as all software projects involve trade-offs, and a lack of systematic methods to do so should lead to worse quality of software and services in the end. Furthermore, we identified a strong correlation between clear functional requirements (\textbf{S5d}) and systematic QR trade-offs (\textbf{S5e}) ($\rho=0.59$). This might, at least in part, be understood by the observation that clear requirements are a prerequisite for systematic trade-offs -- it is impossible to do systematic trade-offs with undefined qualities.

\begin{table}[ht!]
\centering
\caption{Ranking of ISO/IEC 25010 qualities based on the survey and results from the keyword profiling. EN denotes keywords searched for in English only.}
\label{tab:keywords}
\begin{tabular}{l|c|c|c|}
\cline{2-4}
                                             & \multicolumn{2}{c|}{Survey} & Docs     \\ \hline
\multicolumn{1}{|c|}{Qualities}    & \#Mentions       & \#Prio 1s       & \#GovAgs \\ \hline
\multicolumn{1}{|l|}{functional suitability} & 54               & 29              & 1        \\ \hline
\multicolumn{1}{|l|}{performance efficiency} & 11               & 0               & 19       \\ \hline
\multicolumn{1}{|l|}{compatibility}          & 4                & 0               & 10       \\ \hline
\multicolumn{1}{|l|}{usability}              & 34               & 9               & 16       \\ \hline
\multicolumn{1}{|l|}{reliability}            & 30               & 4               & 1 (EN)       \\ \hline
\multicolumn{1}{|l|}{security}               & 46               & 12              & 20\footnotemark      \\ \hline
\multicolumn{1}{|l|}{- confidentiality}      & N/A              & N/A             & 18       \\ \hline
\multicolumn{1}{|l|}{- integrity}            & N/A              & N/A             & 3        \\ \hline
\multicolumn{1}{|l|}{- availability}         & N/A              & N/A             & 4 (EN)        \\ \hline
\multicolumn{1}{|l|}{maintainability}        & 16               & 1               & 4        \\ \hline
\multicolumn{1}{|l|}{portability}            & 2                & 0               & 7        \\ \hline
\end{tabular}
\end{table}
\footnotetext{In the documents, security was not searched for directly, and the number given is the union of the documents where confidentiality, integrity, and availability occurred.}

\section{Conclusion} \label{sec:conc}
Societal digitalization is happening at a rapid pace worldwide, and Sweden is no exception. Digital solutions enable several benefits, but also introduce novel risks as shown by the IT scandal that surfaced in Sweden during 2017. To reap the benefits of digital solutions, and to mitigate the risks involved, considerable understanding of both software and software engineering is essential. Software is continuously scaling in society, and the public sector must ensure that its software maturity matches the new era.

We conducted a census of Swedish Government Agencies (GovAgs) to overview the extent of internal software development projects -- as an indicator of the digitalization of society. Ninety-three GovAgs (39.2\%) confirmed conducting software development (\textbf{RQ1}). While most such GovAgs have developers in the magnitude of tens, several GovAgs have hundreds of development resources. Swedish GovAgs often complement the employed developers with roughly the same number of contractors. However, our survey suggests that relying heavily on contractors is correlated with high technical debt and negatively correlated with coordinated development efforts with other GovAgs -- thus we recommend GovAgs to maintain sufficient software know-how in-house.

GovAgs must make strategic decisions on software sourcing, e.g., whether to develop software internally or acquire it externally using public procurement (\textbf{RQ2}). Most GovAgs focus on in-house development (60.8\%), but outsourcing of development is widespread -- however rarely to organizations outside of Sweden. If possible, a majority of the GovAgs aim to purchase software solutions off-the-shelf (58.1\%). Furthermore, few GovAgs routinely develop software under open source licenses (8.1\%), despite recent calls that software development funded by tax money should be publicly available. On the other hand, somewhat hypocritically, 41.9\% of the GovAgs strive to integrate available OSS into their own solutions. Open innovation initiatives require domain knowledge, and we recommend GovAgs to find inspiration in successful software ecosystems in the private sector. 
Moreover, GovAgs spearheading openness rather than lagging behind would perfectly match OECD's promises of transparency and trustworthiness through digital governments. Opposed to private companies struggling with making profits on the market, Govags are not competitors -- thus it seems like a real opportunity for them to openly collaborate.

Software development in the public sector offers a different context than the market-driven development typically studied by the software engineering community. We provide insights into state-of-practice development in Swedish GovAgs and the prevalence of a selection of software engineering best practices (\textbf{RQ3}). In line with previous work, we conclude that public sector development resembles its private sector counterpart. A majority of the GovAgs develop software iteratively using modern development tools and several best practices are implemented, e.g., software processes, requirements engineering, close communication with end-users. 

Software quality requirements are an acknowledged development challenge that inevitably introduce trade-offs. Based on the qualities defined by ISO/IEC 25010~\cite{international_organization_for_standardization_systems_2011}, our survey shows that GovAgs, in line with private sector companies, tend to prioritize functional suitability. Several GovAgs instead consider security to be the most important (\textbf{RQ4}), and also usability and reliability are frequently highlighted as important qualities. Performance efficiency is rarely a primary concern, but related keywords often occur in the GovAgs' System Requirements Specifications (SRS). More importantly, few GovAgs use systematic approaches to prioritize different qualities. Based on the lack of systematic approaches, combined with indications of poor alignment between strategic goals and operationalized reality, we recommend GovAgs to focus process improvements on software qualities -- to ensure future trust in the software that drive the digitalization of Swedish society.

This paper summarizes the first step in a larger ambition to study software development during the digitalization of the Swedish public sector. Our planned research is needed, as the digital transformation is happening fast and large amounts of tax money are at stake -- the software engineering community ought to contribute its experience and support decision-makers as consultation bodies. 

The Swedish Government has an appointed minister for Housing and Digital Development, but the double duty shows that the question has not been given sufficient weight. There is a silver lining, however, as the Swedish Government announced in the budget bill for 2018 that a new GovAg will be commissioned from September 2018. The new GovAg will shoulder an overall responsibility to coordinate and support the governmental digitalization at large, possibly in line with the Danish Agency for Digitisation -- established already in 2011. We are eagerly awaiting the new GovAg, and hope to contribute our research perspectives for a successful -- and accelerating -- digitalization of the Swedish public sector.

\newpage

\begin{acks}
Thanks go to all respondents. The work is partially supported by a research grant for the ORION project (reference number ~\grantnum{20140218}{20140218}) from The Knowledge Foundation in Sweden and by a grant for the DRISTIG project by the Swedish Civil Contingencies Agency, MSB (agreement no.~\grantnum{2015-6986}{2015-6986}).
\end{acks}

\bibliographystyle{ACM-Reference-Format}
\bibliography{SwedishGov} 

\end{document}